\documentclass[%
 aps,
 jmp,%
 amsmath,amssymb,
  preprint,%
  showpacs,%
]{revtex4-1}
\usepackage{hyperref}%
\usepackage{graphicx}
\usepackage{dcolumn}
\usepackage{bm}
\usepackage{soul}
\usepackage{color}

\usepackage[mathlines]{lineno}
\linenumbers\relax 

\begin{document}

\preprint{APS/123-QED}

\title{Characterization of an INVS Model IV Neutron Counter for High Precision ($\gamma,n$) Cross-Section Measurements}

\author{C.W. Arnold}
 \email{cwarnold@physics.unc.edu}
\author{T.B. Clegg}%
\author{H.J. Karwowski}%
\author{G.C. Rich}%
\author{J.R. Tompkins}%
\affiliation{Department of Physics and Astronomy, University of North Carolina at Chapel Hill, Chapel Hill, NC 27599}
\affiliation{Triangle Universities Nuclear Laboratory (TUNL), Durham, NC 27708}
\homepage{http://www.tunl.duke.edu}
\author{C.R. Howell}%
\affiliation{Department of Physics, Duke University, Durham, NC 27708}
\affiliation{Triangle Universities Nuclear Laboratory (TUNL), Durham, NC 27708}
\homepage{http://www.tunl.duke.edu}
\date{\today}

\begin{abstract}
A neutron counter designed for assay of radioactive materials has been adapted for beam experiments at TUNL.  The cylindrical geometry and 60\% maximum efficiency make it well suited for ($\gamma,n$) cross-section measurements near the neutron emission threshold.  A high precision characterization of the counter has been made using neutrons from several sources.
Using a combination of measurements and simulations, the absolute detection efficiency of the neutron counter was determined to an accuracy of $\pm$ 3\% in the neutron energy range between 0.1 and 1 MeV.   It is shown that this efficiency characterization is generally valid for a wide range of targets.  %

\end{abstract}

\maketitle

\section{\label{sec:level1}Introduction:\protect}

The model IV inventory sample counter (INVS) developed at Los Alamos National Laboratory \cite{Sprinkle93} was designed for fast, non-destructive assay of radioactive materials.  
Specialized inserts for the axial bore of this neutron counter have been made to adapt it for use as the primary neutron detector for in-beam ($\gamma$,n) total cross section measurements.  Development and testing of this counter took place at TUNL using hadron beams in the tandem laboratory and the $\gamma$-ray beam at the High Intensity Gamma-ray Source (HI$\gamma$S)~\cite{higsref}.  


Such measurements require detailed and accurate information about the energy-dependent absolute neutron detection efficiency of the counter.  Efficiency here is generally defined as 
\begin{equation}
\epsilon \equiv \frac{N_{detected}}{N_{emitted}}
\label{eq:see}.
\end{equation}
The efficiency measurements were made using four different sources, each with a precisely known neutron emission rate.  %
First, a $^{252}$Cf source, calibrated by the National Institute of Standards and Technology (NIST), generated a flux of neutrons known to $\pm$ 4.4\% \cite{Tho09}.
Second,
a coincidence experiment using the $^{2}$H($d,n$)$^{3}$He reaction provided a mono-energetic source of 2.26 MeV neutrons with flux known to $\pm$ 10\%, and gave insight into the thermalization time of neutrons in the INVS.
Third, 
an investigation of the $^{7}$Li($p,n$)$^{7}$Be reaction produced $<$1 MeV neutron sources with fluxes known to $\pm$ 6.6\%.  
Finally, 
the $^{2}$H($\gamma,n$)$^{1}$H reaction was used to produce tunable sources of monoenergetic neutrons (0.1 $\leq$ E$_{n}$ $\leq$ 1.0 MeV) with fluxes known to $\pm$ 3\% accuracy. 
A comparison of all experimental data with simulations demonstrates varying levels of agreement. 


The following sections describe 
the detector geometry (Sect. \ref{sec:DD}) followed by details for each experimental setup (Sect. \ref{sec:EXP}) including discussions of backgrounds, calculations, measurement uncertainties and results. Section \ref{sec:DISCUSS} contains a discussion of Monte Carlo simulations used for comparison with experimental results.  Finally, Sect. \ref{sec:SUM} offers a summary of the results and a discussion of future applications of the INVS counter for cross-section measurements.

\section{\label{sec:DD}Detector Description:\protect}
\begin{figure}
\includegraphics[width=0.5\textwidth]{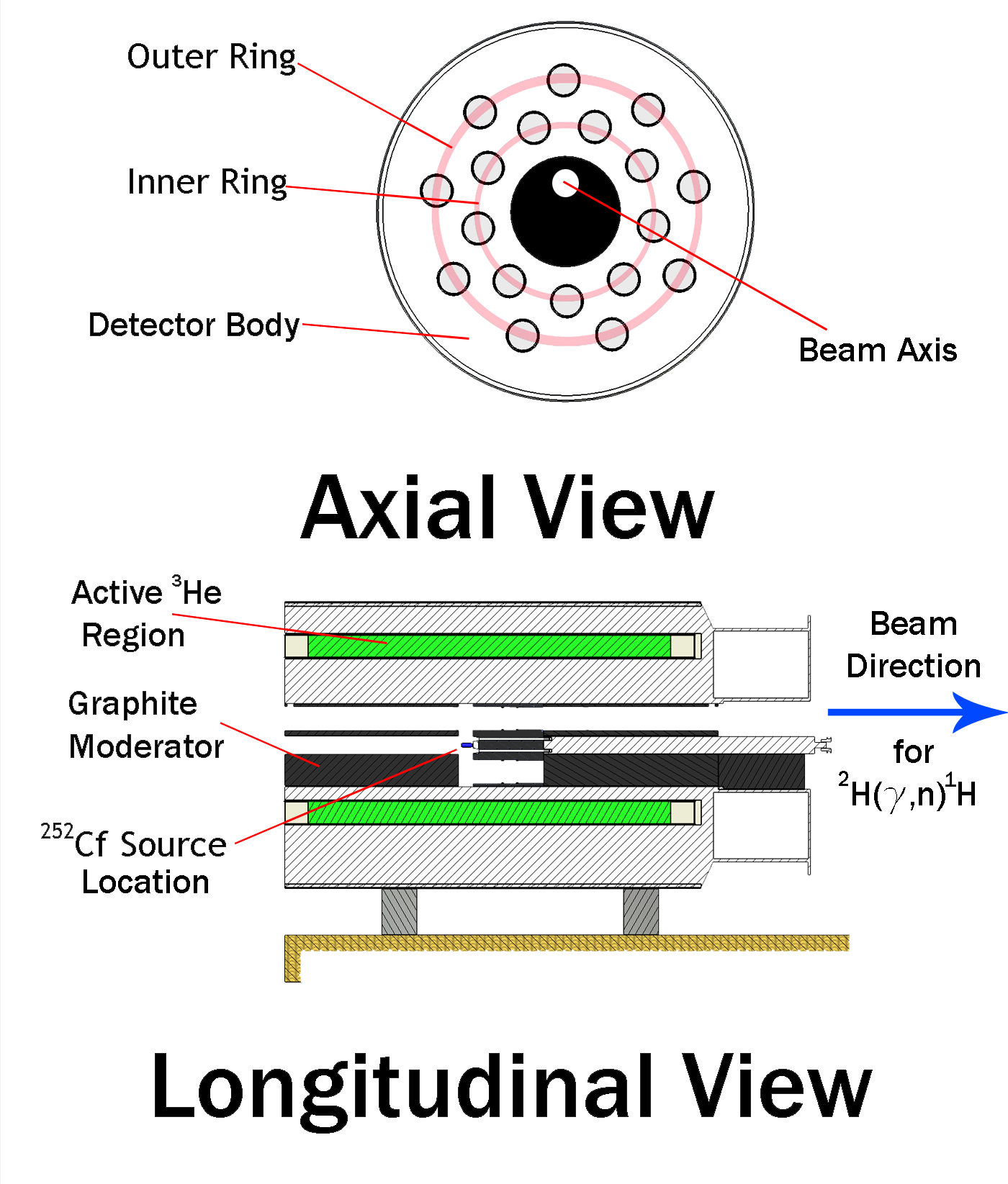}
\caption{\label{Fig:detview}(Color Online) Front and side cut-away cross-sectional views of the Model IV INVS counter described in the text.  The arrangement of inter-cavity moderator corresponds to the experimental geometries either for the $^{252}$Cf source measurement, or for the $^{2}$H($\gamma, n$)$^{1}$H experiment.}
\end{figure}

The active detection elements in the INVS counter are 18 tubular proportional counters, each containing 6 atm. of $^{3}$He.  The tubes are arranged in two concentric rings at radii 7.24 cm and 10.60 cm each containing nine equally spaced detectors (see Fig. \ref{Fig:detview}). The detectors are embedded in a cylindrical polyethylene body 46.2 cm long and 30.5 cm in diameter which serves as a neutron moderator. The active length of the $^{3}$He gas within the tubes is 39.4 cm~\cite{Sprinkle93}.  The detector body has an 8.9 cm diameter axial cavity designed to contain the neutron source. Throughout this manuscript the term \emph{longitudinal center} refers to the center of the detector
with respect to the length of the detector body, and is distinguished from the term \emph{axial center} which refers to the axis of the detector.

In experiments with beam, the irradiated target is the source of neutrons.  During all but one of the experiments described here, the neutron-emitting source was located inside the detector cavity, usually near the longitudinal center.  
The cavity was partially filled with additional neutron moderator (often graphite and/or polyethylene) to increase the detection efficiency.  The additional moderator was arranged so that the beam could pass through the detector without intercepting moderator material.

Thermalization of the neutrons within the detector body increases the probability for initiating the $^{3}$He($n,p$)$^{3}$H reaction within the embedded tubes.  An energy of 763.7 keV, shared between the outgoing proton and triton, is released from each reaction. Most of the kinetic energy is lost to ionization of $^3$He, which is detected as an electrical pulse on the central electrode of each tube which is biased to $+$1780 V.  A fixed threshold effectively discriminates against low-pulse-height signals generated by $\gamma$-rays and electronic noise.  Signals above the threshold generate $\sim$50 ns wide TTL pulses using on-board electronics. Onboard signal-processing electronics within the detector produce three TTL logic output signals; the inner ring ($I$); the outer ring ($O$); and the logical OR of the $I$ and $O$ pulses ($T$).  Whenever one or more tubes in the inner (outer) ring detect a neutron, a pulse is generated on the $I$ ($O$) output.  For neutrons with energies less than about 2 MeV, the I/O ratio can provide a reasonable determination of the mean of the energy distribution of the detected neutrons. 

\begin{figure}
\includegraphics[width=0.5\textwidth]{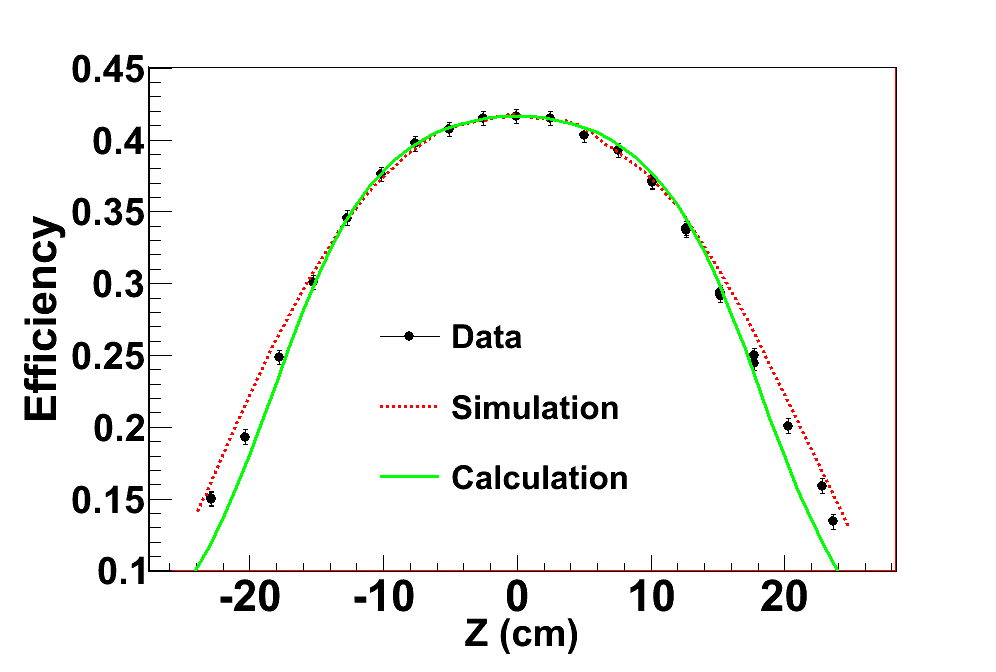}
\caption{\label{FIG:effz}(Color Online) Efficiency vs. Z-axis position for an open detector geometry.}
\end{figure}

\section{\label{sec:EXP}Experiments:\protect}
In this section, the experimental setup and techniques used to measure the efficiency of the counter are described, and the results of the each measurement are presented.  Discussion of the results will be presented in the next section.  Four different sources of neutrons were used to cover an energy range from about 0.1 to 10 MeV.
\subsection{\label{sec:CF252}$^{252}$Cf}
Californium-252 is a standard calibration source for neutron detectors.  
The effective half life of $^{252}$Cf is 2.645 years and is due to alpha particle emission and spontaneous fission which produces a neutron yield of 2.314 $\times$ 10$^6$ neutrons/s/$\mu$g \cite{Martin99}.  The energy spectrum of neutrons from $^{252}$Cf is well known \cite{Smith57}.
A calibrated $^{252}$Cf source provides a single measurement of efficiency representing the response of the detector to a broad spectrum of neutron energies.  

The $^{252}$Cf source used consisted of 3.15 ng of active powdered material encapsulated within a small aluminum pellet.  The source was suspended on the axis of the detector approximately 3.4 cm from longitudinal center.  Graphite moderator filled most of the volume of the cavity.  A table of neutron activity versus date was provided by NIST \cite{Tho09}.  The experimentally determined efficiency for this configuration (shown in Fig.~\ref{Fig:detview}) was 40.5 $\pm$ 1.8 \%. 
The experimental I/O ratio was 1.516 $\pm$ 0.004. 

The dependence of the detection efficiency on the position of the source within the central cavity was determined by making measurements with the source placed at different positions within the central bore.  Measurements on the central axis were made along the entire length of the detector. The detection efficiency has a maximum value at the longitudinal center and drops off smoothly as the source is moved in either direction away from the center along the detector axis.  The shape of the position dependency is a purely geometric acceptance effect and can be approximated analytically for point sources with isotropic neutron emission.  The measured detection efficiency as a function of the source position along the central axis of the counter is shown in Fig.~\ref{FIG:effz} and in comparison to simulated and calculated efficiencies.  

The relative efficiency is directly proportional to the angular acceptance of the counter as a function of z, which is given by the equation below for an isotropic point source of neutrons. 

\begin{eqnarray}
 \epsilon \propto d\Omega \approx~4\pi~~~~~~~~~~~~~~~~~~~~~~~~~~~~~~~~~~~~~~~~~~\nonumber\\
  - 2\pi\times\left[\left(1-\frac{L/2 - z}{\sqrt{(L/2-z)^{2} + r^{2}}}\right)\right.~~~~~~~~~~~~\nonumber\\
  ~~~~~~~~~~~\left.+\left(1-\frac{L/2 + z}{\sqrt{(L/2+z)^{2} + r^{2}}}\right)\right]
\end{eqnarray}

Here, r is the radius of the opening at the end of the detector, L is the active length of the $^{3}$He gas, and z = 0 at L/2.  This function is maximum when z = 0.  %
For an open cavity geometry, the change in the efficiency over the length of an 8-cm long sample, centered on the axis at the longitudinal center is approximately 1\%. For a geometry like the one shown in Fig.~\ref{Fig:detview} the changes in efficiency are negilgibly small over a length of nearly 20 cm centered on the longitudinal center. 
For sources located off its central axis, the detection efficiency changes by less than 0.5\% . The radial dependence of the efficiency is also mostly a geometric acceptance effect.

\subsection{\label{sec:DDN}$^{2}$H($d,n$)$^{3}$He}

\begin{figure}
\includegraphics[width=0.5\textwidth]{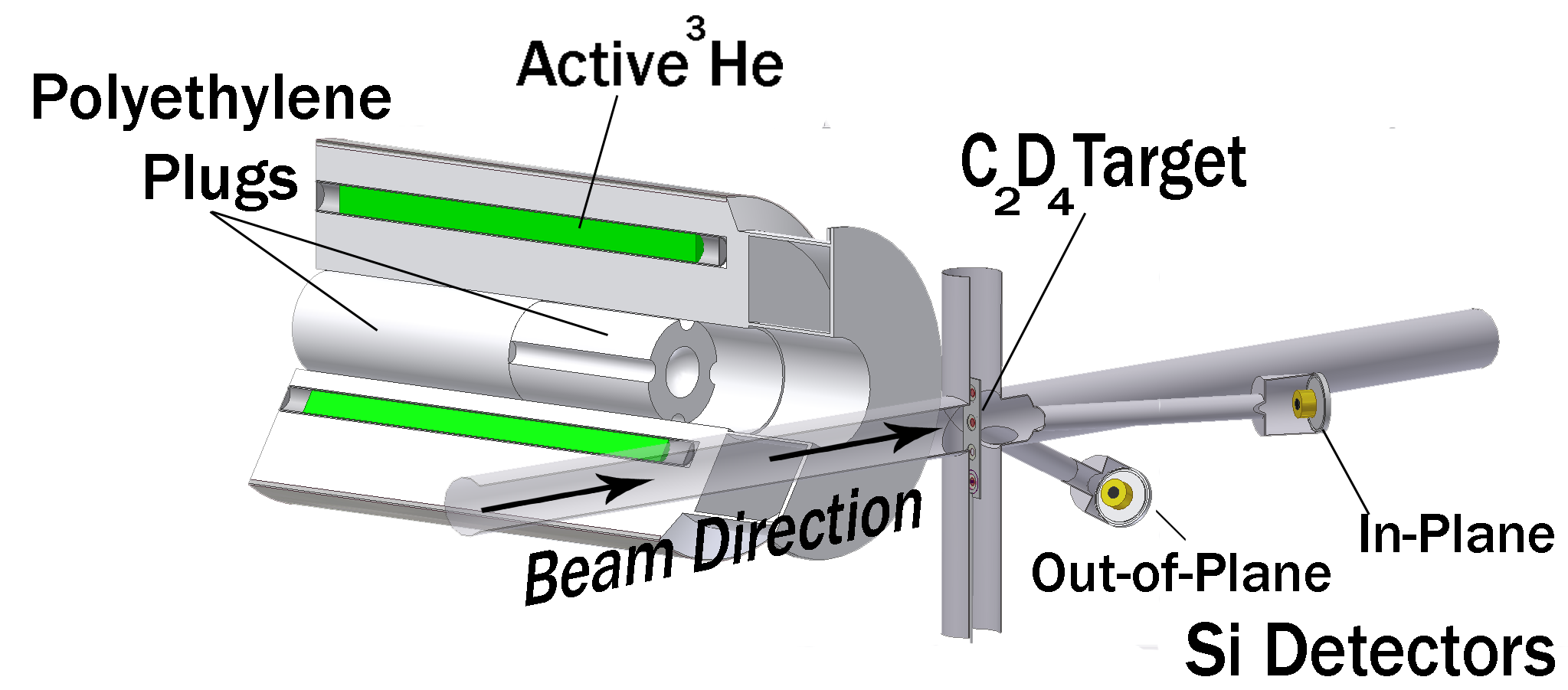}
\caption{\label{Fig:ddngeom}(Color Online) Schematic diagram of the experimental setup for the efficiency measurements made using the $^{2}$H($d,n$)$^{3}$He reaction at the tandem accelerator facility.}
\end{figure}

The $^{2}$H($d,n$)$^{3}$He reaction was used to measure the efficiency for monoenergetic 2.26 MeV neutrons.  The associated particle technique was used with the recoil $^3$He nucleus detected in a siliFcon surface barrier detector inside an evacuated chamber. A schematic diagram of the experiment setup is shown in Fig. \ref{Fig:ddngeom}.  The neutron counter was positioned so that its central axis coincided with the symmetry axis of the cone of neutrons associated with the $^3$He particles detected in the silicon detector on the opposite side of the incident beam axis. The distance from the longitudinal center of the counter to the deuterium target was set so that the diameter of this neutron cone was smaller than the diameter of the central cavity through the detector. The energy of the incident deuteron beam and the detection angle of the silicon detector were set to produce 2.26-MeV neutrons emitted along the central axis of the counter.  With this method the efficiency is computed as 
\begin{equation}
\epsilon = \frac{N_{n}}{N_{^{3}He}}
\end{equation}
where N$_{^{3}He}$ is the total number of detected $^{3}$He-particles and $N_{n}$ is the total number of neutrons detected in coincidence with the detected $^3$He particles.  This equation takes the detection efficiency of the silicon detector to be unity.
The deuterium targets used in these measurements were $\sim$ 100$\mu$g/cm$^{2}$ thick deuterated polyethylene (C$_{2}$D$_{4}$) evaporated onto a 10 $\mu$g/cm$^2$ thick carbon foil.  The deuteron beam energy incident on the foil was 2.0 MeV, and the average beam current on the C$_{2}$D$_{4}$ foil was $\sim$20 nA. The cross-sectional profile of the deuteron beam at the foil was circular with a diameter of approximately 0.5 cm. 
%
%
%
Each of the two silicon detectors (one in-plane and one out-of-plane) had a solid angle acceptance d$\Omega$ = $\pi$/60 sr, and each was located at a scattering angle of $\theta_{lab}$ = 26.50$^{\circ}$.  Neutrons associated with detection of $^{3}$He in the in-plane Si detector exited the target at $\theta_{lab}$ = 117.1$^{\circ}$ along the central axis of the neutron counter. The rear half of the central cavity was plugged with polyethylene to scatter neutrons traveling through the central cavity into the body of the counter.   The out-of-plane Si detector was used to measure the rate of accidental coincidences.

\begin{figure}
\includegraphics[width=0.5\textwidth]{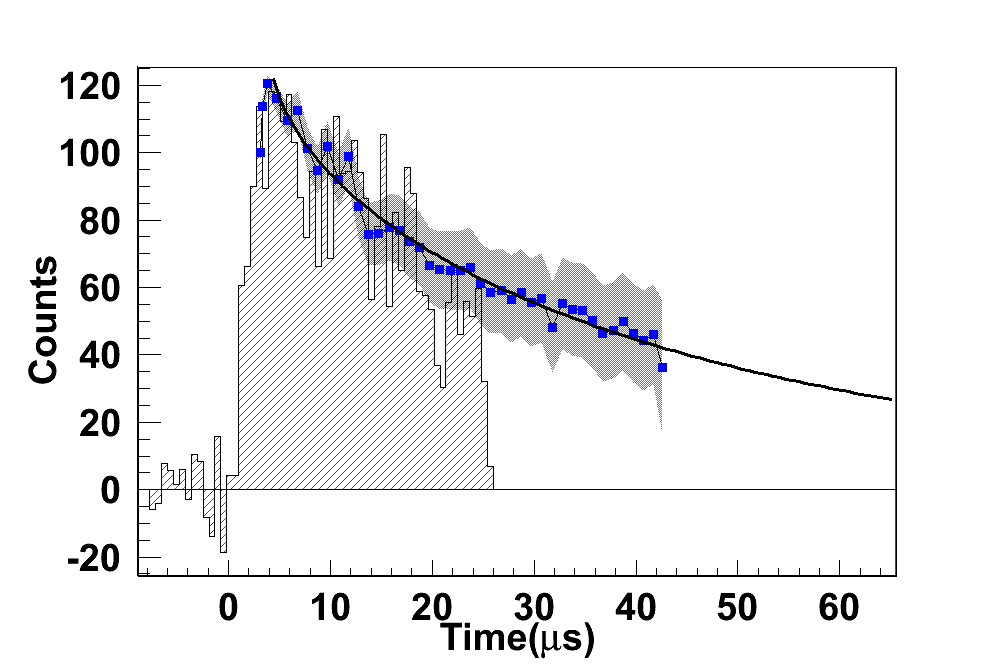}
\caption{\label{Fig:TACSPEC}(Color Online) The histogram is a background subtracted TAC spectrum from the $^{2}$H($d,n$)$^{3}$He experiment. A simulated TAC spectrum (blue points) is shown in comparison to the experimental histogram.  The solid curve is a fit to the simulated spectrum and predicts that most neutron detection occurs outside the 22.5 $\mu$s experimental window.}
\end{figure}

Efficiency and thermalization time were deduced simultaneously using a time-to-amplitude converter (TAC) %
which recorded the time between a charged particle detection in the silicon detector and a neutron detection in the INVS.  
A threshold setting effectively discriminated against deuteron elastic scattering events.  The effective TAC range was 22.5 $\mu$s.  The TAC was calibrated using a pulser which started and stopped the TAC with known delay.  Because the INVS is a thermalization counter, detection efficiency is time-dependent on a microsecond time-scale. A peak in the TAC spectrum at $\sim$3 $\mu$s suggests a source of delay exists caused by charge collection and signal processing in the INVS counter.  This delay reduces the effective TAC range to 22.5 $\mu$s from 25.5 $\mu$s, which is where the experimental TAC spectrum ends. The present result for neutron detection efficiency over a 22.5 $\mu$s range is 11.0 $\pm$ 1.1\%.  

\subsection{\label{sec:LIPN}$^{7}$Li($p,n$)$^{7}$Be}
The $^{7}$Li($p,n$)$^{7}$Be reaction was used to measure the energy-dependent detection efficiency over an energy range that overlaps with that covered by the $^{2}$H($\gamma,n$)$^{1}$H source reaction below about 0.7 MeV and to provide data for a neutron source with the intensity distribution peaked at forward angles relative to the central detector axis \cite{Bur74, Bur72}. The cross section for $^{7}$Li($p,n$)$^{7}$Be reaction is large and has been accurately measured \cite{Gib59} making it a good neutron source for calibrating the efficiency of detectors at low energies \cite{Sek76}. 

\subsubsection{\label{sec:TPREPLi7}Experimental Setup}
\begin{figure}
\includegraphics[width=0.5\textwidth]{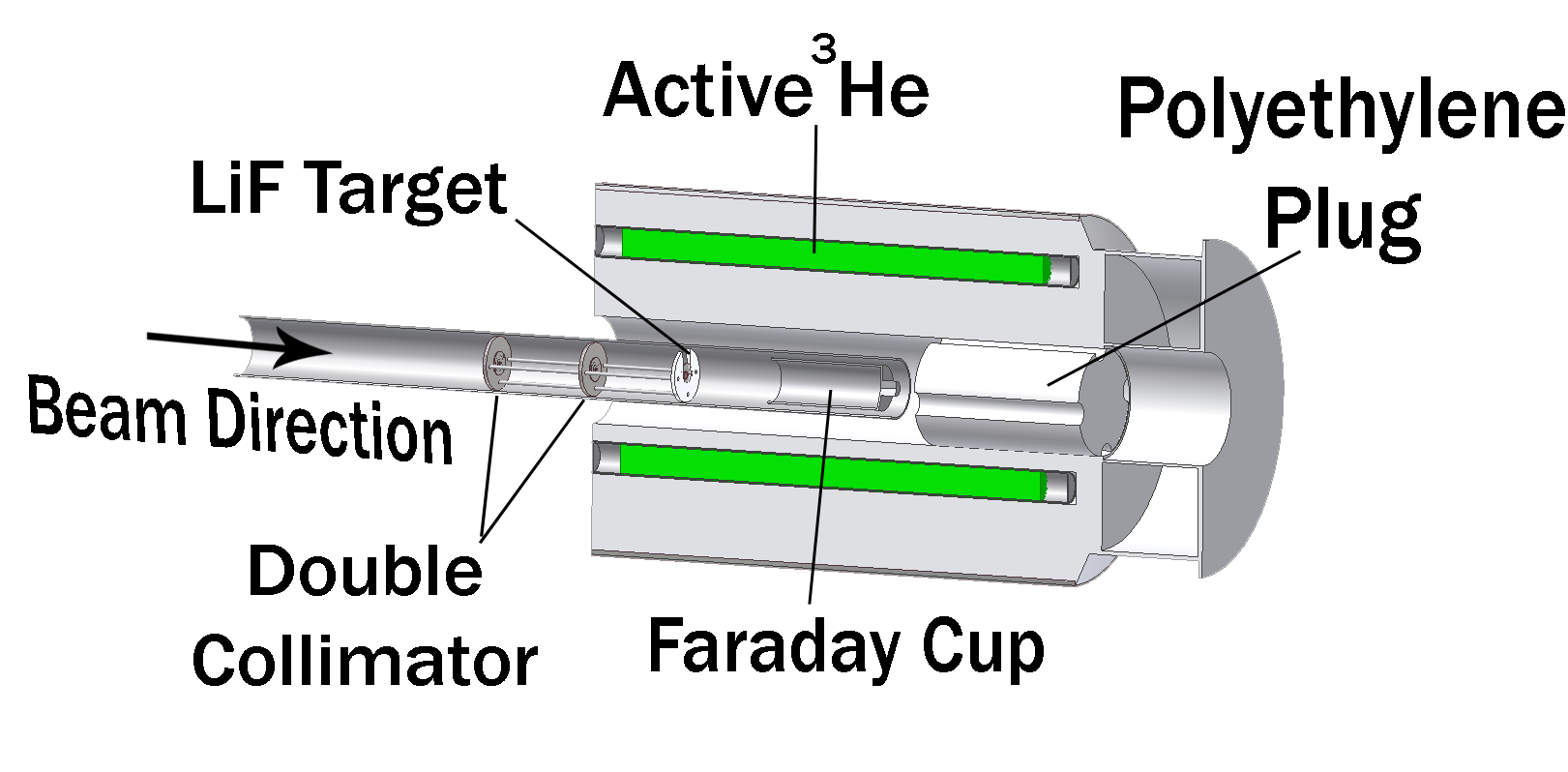}
\caption{\label{Fig:Li7geom}(Color Online) Schematic diagram of the experimental setup for the efficiency measurements made using the $^{7}$Li($p,n$)$^{7}$Be reaction at the tandem accelerator facility.}
\end{figure}

The experimental arrangement is shown in Fig.~\ref{Fig:Li7geom}.  The proton beam was tuned through a double collimator set onto the LiF neutron production target. The cross-sectional profile of the beam on target was circular with a diameter of 5 mm, and the average beam current on target was 100 nA. The energies of the proton beams incident on the LiF target were between 1.88 and 2.46 MeV.  The neutron production target was comprised of 39.8 $\mu$g/cm$^{2}$ of LiF evaporated onto a 8.3 $\mu$g/cm$^{2}$ thick carbon backing. Targets were located on the axis of the INVS counter inside an evacuated beam pipe at 14.2 cm from the longitudunal center.  The transmitted proton beam was collected in a voltage-suppressed Faraday cup at the end of the beam pipe.  A polyethylene plug was placed just beyond the end of the beam pipe to increase detection efficiency. Backgrounds were measured by putting beam through both an empty target ring identical to the one that supported the LiF target, and a target ring that supported only a carbon backing. In total, beam-induced and environmental backgrounds amounted to $\leq$ 0.1\% of real counts.

\subsubsection{\label{sec:CALCLi7}Results}
The detection efficiency as a function of proton energy was calculated using
\begin{equation}
\epsilon(E_{p}) = \frac{N_{n}}{N_{p}N_{t}\sigma(E_{p})}
\end{equation}
where N$_{n}$ is the total number of neutrons detected, $N_{p}$ is the number of protons collected in the Faraday cup, $N_{t}$ is the number of target nuclei per unit area, and $\sigma(E_{p})$ is the total cross-section of the $^{7}$Li($p,n$)$^{7}$Be reaction at proton energy E$_{p}$. 

\begin{figure}
\includegraphics[width=0.5\textwidth]{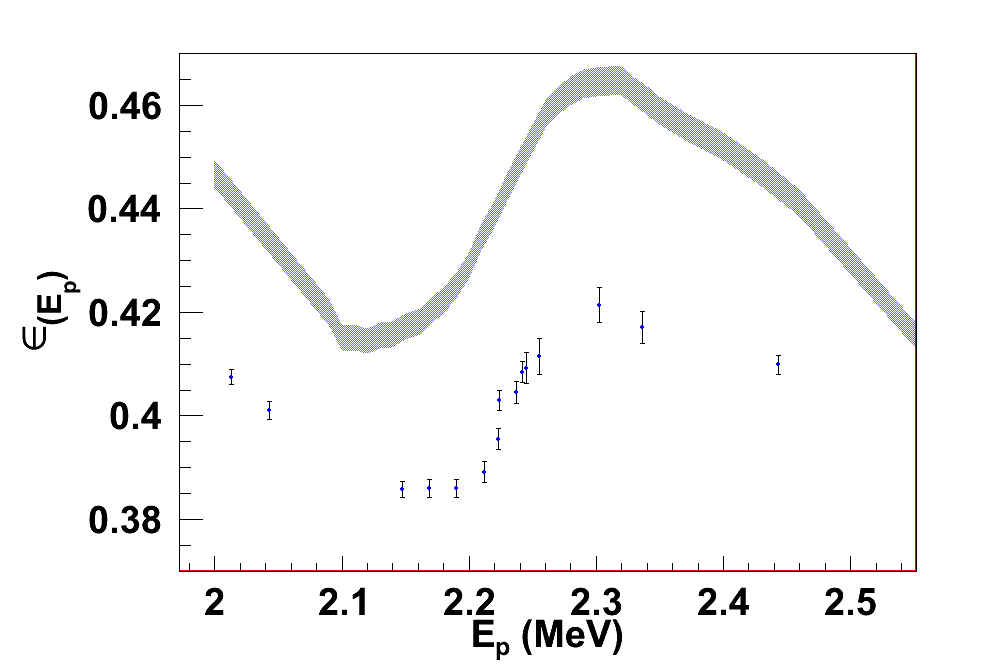}
\caption{\label{Fig:Li7_total}(Color Online) For the $^{7}$Li($p,n$)$^{7}$Be reaction, a comparison of $\epsilon(E_{p})$ as determined by experiment (points) and by simulation (band). Simulations, discussed in Sect. \ref{sec:DISCUSS}, reproduce the shape of the efficiency well. }  
\end{figure}

The data (see Fig.~\ref{Fig:Li7_total}) display a relative minimum in efficiency near E$_{p}$ = 2.13 MeV followed by a relative maximum near E$_{p}$ = 2.32 MeV.  These shifts in efficiency coincide with rapid changes in the angular distribution of neutrons. Though statistical uncertainties were very small, systematic uncertainties for target thickness and cross-section contributed 3.5\% and 5\%, respectively, resulting in an overall systematic uncertainty of 6.6\%.

\subsection{\label{sec:DGN}$^{2}$H($\gamma, n$)$^{1}$H}

\begin{figure*}
\includegraphics[width=0.95\textwidth]{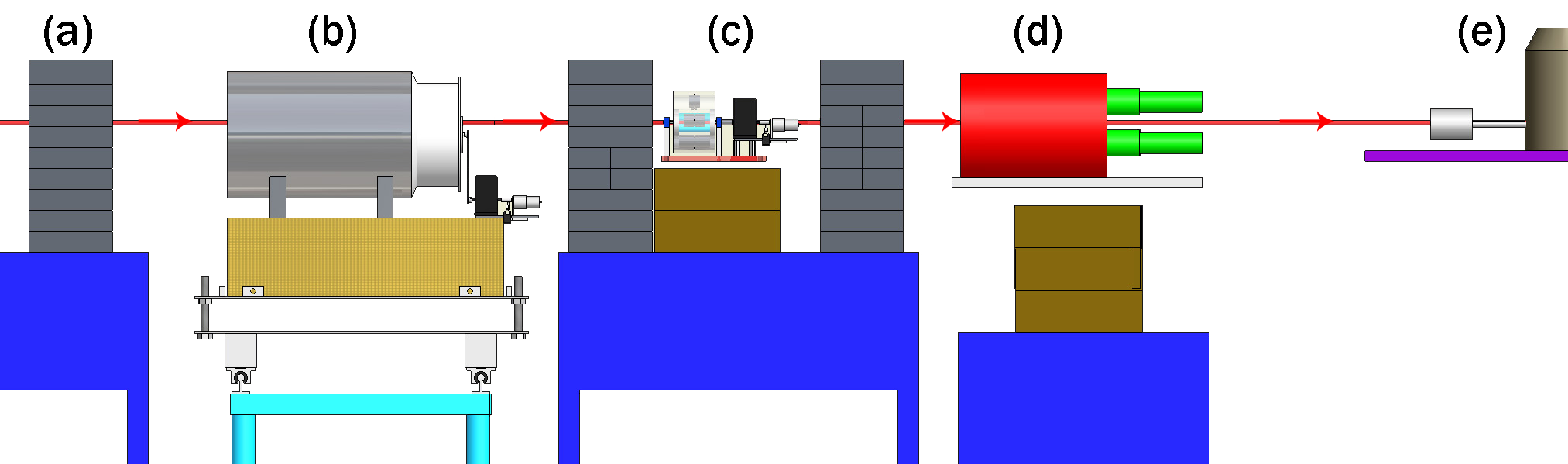}
\caption{\label{Fig:HIGSSchm}(Color Online) Schematic of the experimental setup for measurements made using the $^{2}$H($\gamma, n$)$^{1}$H reaction at the HI$\gamma$S facility.  After collimation, the $\gamma$-ray beam passes through scintillation paddles (not shown) and into the target room.  The 1.2 cm diameter $\gamma$-ray beam passes through the following elements: (a) "clean-up" collimator wall; (b) the D$_2$O target located near the longitudinal center of the neutron counter; (c) machined lead attenuators located between lead collimator walls; (d) a NaI detector;(e) an HPGe detector. }
\end{figure*}

The $^{2}$H($\gamma, n$)$^{1}$H measurement was unique among the experiments described here in that it produced nearly monoenergetic neutrons with very small flux uncertainties.  Several efficiency measurements were made that highlighted the energy-dependent response of the detector.  A schematic of the experimental setup is shown in Fig \ref{Fig:HIGSSchm}.

\subsubsection{\label{sec:TPREPDGN}Experimental Setup}
The target consisted of approximately 3.2 g of 99.8\% enriched D$_{2}$O sealed inside a thin-walled polyethylene tube that was 7.62 cm long.  The target was located at longitudinal center, but 2.9 cm above the axial center, and occupied one of four 2.54 cm diameter azimuthal holes in a rotatable graphite cylinder. The other holes were available for other targets.  A graphite target and an empty hole were alternately rotated into the beam during experiments to determine beam-induced backgrounds. The samples were remotely rotated into position using a four-position Geneva mechanism which assured reproducible alignment on the $\gamma$-ray beam axis. Most of the data were collected for $\gamma$-ray beams with energies between 2.48 and 4.10 MeV.

Gamma-ray production by inverse-Compton scattering at HI$\gamma$S is well documented \cite{Lit97, Sun2009, higsref}.  The $\gamma$-ray beams used in this experiment were collimated to 1.2 cm in diameter.  
The size of the beam, and its alignment with the target was confirmed using a $\gamma$-ray beam imaging system \cite{Sun08}.

\begin{figure}
\includegraphics[width=0.5\textwidth]{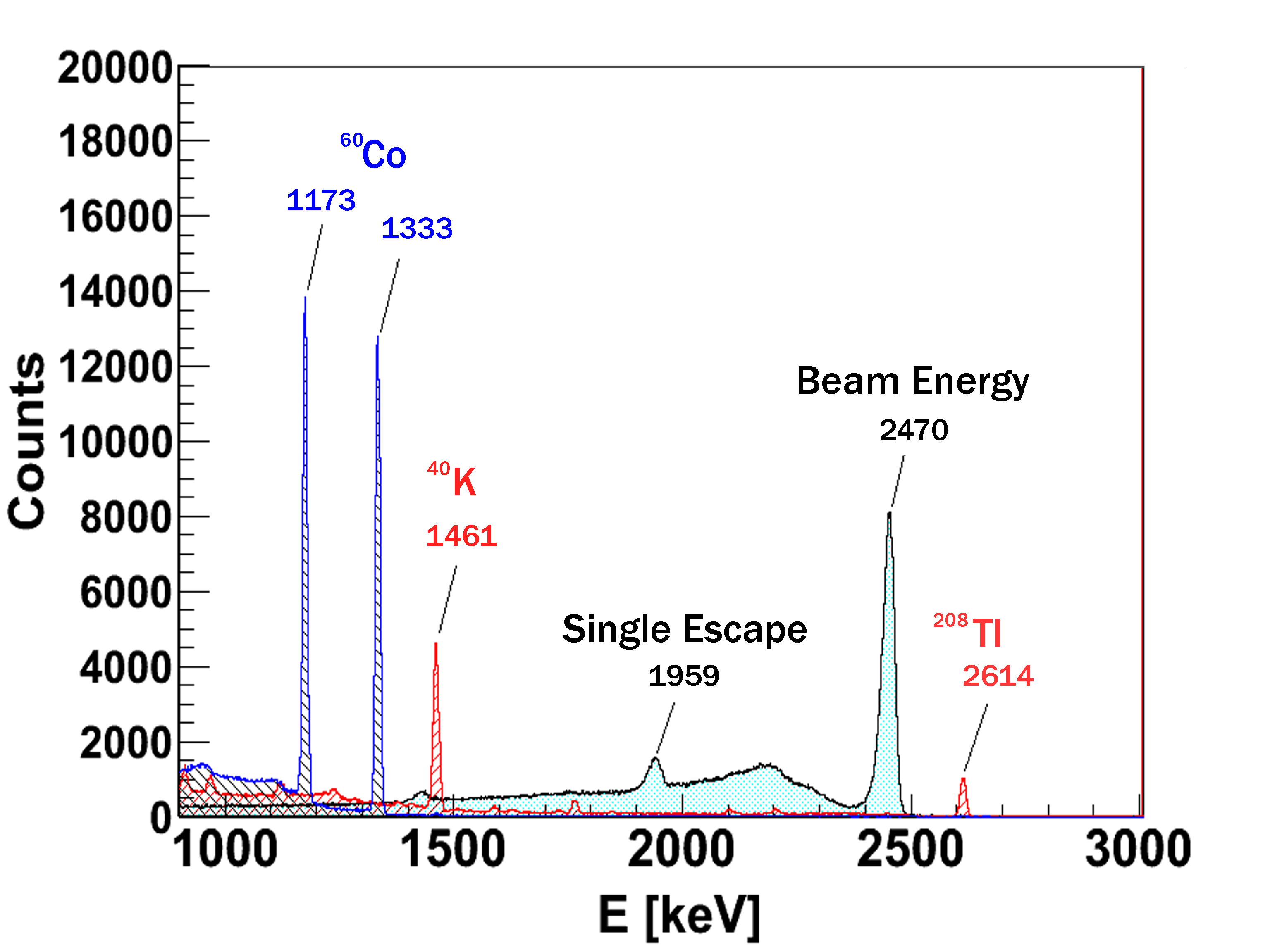}
\caption{\label{Fig:g_stats}(Color Online) A HPGe spectrum for 2.470 MeV $\gamma$-ray beam with dE/E = 1\%.  Spectra of calibration $\gamma$-rays from $^{60}$Co, $^{40}$K, and $^{208}$Tl are overlaid.}
\end{figure}

The $\gamma$-ray beam energy distribution was determined by a high purity germanium (HPGe) detector located on the $\gamma$-ray beam axis.  A radioactive $^{60}$Co source, and naturally present $^{40}$K and $^{208}$Tl provided energy calibration.  The FWHM of the $\gamma$-ray beam as determined by the HPGe was typically between 1 - 3\%.  A typical $\gamma$-ray spectrum is shown in Fig.~\ref{Fig:g_stats}.  

Relative incident $\gamma$-ray flux was continuously monitored by three scintillating paddles located upstream from the experimental setup.  The absolute $\gamma$-ray fluxes were determined by a cylindrical 25.4 cm $\times$ 35.6 cm NaI detector located behind the active target, on the $\gamma$-ray beam axis.  
For the $\gamma$-ray beam energies for which it was used, the NaI detector had a total integrated efficiency of nearly 100\%; In other words, nearly all $\gamma$-rays will interact within the detector volume and deposit energy.  A threshold setting which ignored signals generated from $\gamma$-rays with energy less than $\sim$0.6 MeV reduced the efficiency to $\sim$97\% for all experimental $\gamma$-ray energies as determined by simulations.  This threshold setting optimized the ratio of total integrated signals to room background.

After passing through the target, the $\gamma$-ray flux was attenuated by machined lead attenuators to eliminate signal pile-up and dead-time effects. This allowed $\gamma$-ray flux on target of $\geq$10$^{6} \gamma$/s, and flux at the face of the detector of $\leq$10$^{4} \gamma$/s. Determination of the absolute beam flux from the signals measured in the NaI detector requires precise quantitative information about the effective attenuation of the beam by the lead attenuators at each $\gamma$-ray energy.  The effective $\gamma$-ray attenuation, was determined for each attenuator, at several $\gamma$-ray energies, using the scintillating paddle system for flux normalization.

\subsubsection{\label{sec:BKDGN}Background}
Two sources of neutron backgrounds existed in this experiment. One source is classified as environmental backgrounds.  These can be caused by either cosmic-ray neutron production or natural radioactive sources in the vicinity of the INVS counter. These sources generated 0.2 count/s per $^{3}$He tube for a total of 3.6 counts/s.  The second background source arose from $\gamma$-ray-beam-induced events.  Gamma-rays that scatter from the target can deposit enough energy to register a signal above the threshold setting.  This type of background was measured by bombarding a graphite target with the $\gamma$-ray beam.  
Gamma-ray beam-induced backgrounds amounted to approximately 4.8 counts/10$^{6}$ $\gamma$-rays on target, which was typically $\leq$ 1\% of real counts from the heavy water target, and were taken into account in the data analysis.

\subsubsection{\label{sec:CALCDGN}Results}
For the experimental setup used in these measurements the detection efficiency for neutrons from $\gamma$-rays on a heavy water sample can be explicitly calculated from
\begin{equation}
\epsilon_{n}(E_{\gamma}) = \frac{N_{n}\chi(E_{\gamma})\epsilon_{\gamma}(E_{\gamma})}{fN_{\gamma}N_{t}\sigma(E_{\gamma})}
\label{eq:eff},
\end{equation}
where N$_{n}$ is the number of neutrons detected, $\chi$(E$_{\gamma}$) is the measured $\gamma$-ray attenuation by the lead attenuator at $\gamma$-ray energy E$_{\gamma}$, $\epsilon_{\gamma}$( E$_{\gamma}$) is the efficiency of the NaI detector for $\gamma$-rays with energy E$_{\gamma}$, f is the thick target correction factor (described below),
$N_{\gamma}$ is the number of $\gamma$-rays detected, $N_{t}$ is the number of target nuclei per unit area and $\sigma(E_{\gamma})$ is the total cross-section of the $^{2}$H($\gamma, n$)$^{1}$H reaction at E$_{\gamma}$ \cite{Sch05}. The author of Ref.~\cite{Sch05} calculated the $^{2}$H($\gamma,n$)$^{1}$H total cross-sections with several widely-used N-N potential models, all of which were indistinguishable to within 1\%, irrespective of the model used. In addition, this cross-section agrees with the world data which have uncertainties between 3 and 6\% \cite{Bir85, Moreh89, DeGraeve92, Hara2003}.  %

The f factor in Eqn.~\ref{eq:eff} accounts for the flux loss due to interactions with atomic electrons as the $\gamma$-rays propagate through the heavy water target.  This factor is calculated as,
\begin{equation}
f = \left(\frac{1}{e^{-\mu_{w} t}}\right)\left(\frac{(1-e^{-\mu_{w}t})}{\mu_{w}t}\right)
,
\end{equation}
using NIST attenuation coefficients for heavy water $\mu_{w}$~\cite{Hub89} and the mass thickness $t$ of the target. 

For photodisintegration of the deuteron $E_{\gamma}$ and $E_{n}$ are related, in units of MeV, by
\begin{equation}
E_{n}=\frac{E_{\gamma} - 2.225}{2.001}
\label{eq:n_energy}
\end{equation}
where -2.225 is the Q-value for the reaction, and the factor of 2.001 comes from energy sharing between the outgoing proton and neutron. 
Cross-sections for energies 2.48 MeV $\leq$ E$_{\gamma}$ $\leq$ 4.10 MeV were used to ensure $\pm$ 1\% accuracy. 

The total combined statistical uncertainty in the efficiency measurements made using neutrons from deuteron photodisintegration was $<$ 2\%. The main source of statistical error was in the attentuation measurements of the lead absorbers.  The statistical accuracy of these measurements was determined by the counts in the scintillator paddle system used for relative flux normalization between the absorbers. A minimum of $\pm$ 1\% statistical uncertainty was achieved for most energies.  The statistical uncertainties in the $\gamma$-ray beam flux measurements and the neutron counting for the heavy water target were typically $\leq$ 0.5\%.  

The total systematic uncertainty was $<$ 3\% and was mainly due to three sources.
The uncertainty in the deuteron photodisintegration cross section was kept to less than 1\% by limiting the energy range of the measurements to 2.48 MeV $\leq$ E$_{\gamma}$ $\leq$ 4.14 MeV.  Uncertainties in the target thickness and cross section contribute 0.5\% and 1\% respectively.

\section{\label{sec:DISCUSS}Discussion and Results:\protect}
\subsection{\label{sec:SIMS}Simulations:\protect}


The Monte-Carlo code \protect{\sc{mcnpx}} was used to simulate all particle interactions within the INVS counter for each experiment.  For all simulations, material densities for the INVS counter were fixed, and standard thermal neutron capture cross-section libraries \cite{Chad06} were used.  
Variable parameters in each simulation were:
\begin{enumerate}
	\item the arrangement of materials inside the cavity;
	\item the location of the neutron emitting source;
	\item the energy and spatial distribution of neutrons.
\end{enumerate}
Absolute detection efficiencies and I/O ratios were extracted from simulations for comparison with each experiment where applicable.

\subsubsection{\label{sec:CFMOD}$^{252}$Cf\protect} 
Efficiency measurements of $^{252}$Cf were made during the course of the $^{2}$H($\gamma, n$)$^{1}$H experiment.  Consequently, the arrangement of materials inside the INVS cavity was identical for both experiments.  The energy distribution of neutrons produced by fissioning $^{252}$Cf was modeled as a Watt fission spectrum which has the form
\begin{equation}
p(E) = Cexp^{(-E/a)}sinh(bE)^{1/2},
\end{equation}
where $a$ and $b$ are parameters given for $^{252}$Cf \cite{mcnpx07}. 

The \protect{\sc{mcnpx}}-simulated efficiency of 39.2\%  agrees with the experimentally determined efficiency of 40.5 $\pm$ 1.8 \%. The simulated I/O ratio of 1.59 falls short of agreement with the experimentally determined 1.516 $\pm$ 0.004 due to a 6\% larger efficiency for measurement in the outer ring (see Fig.~\ref{Fig:AllPlot}). The source of this discrepancy is unclear.

\subsubsection{\label{sec:DDNMOD}$^{2}$H($d,n$)$^{3}$He\protect} 
\begin{figure}
\includegraphics[width=0.5\textwidth]{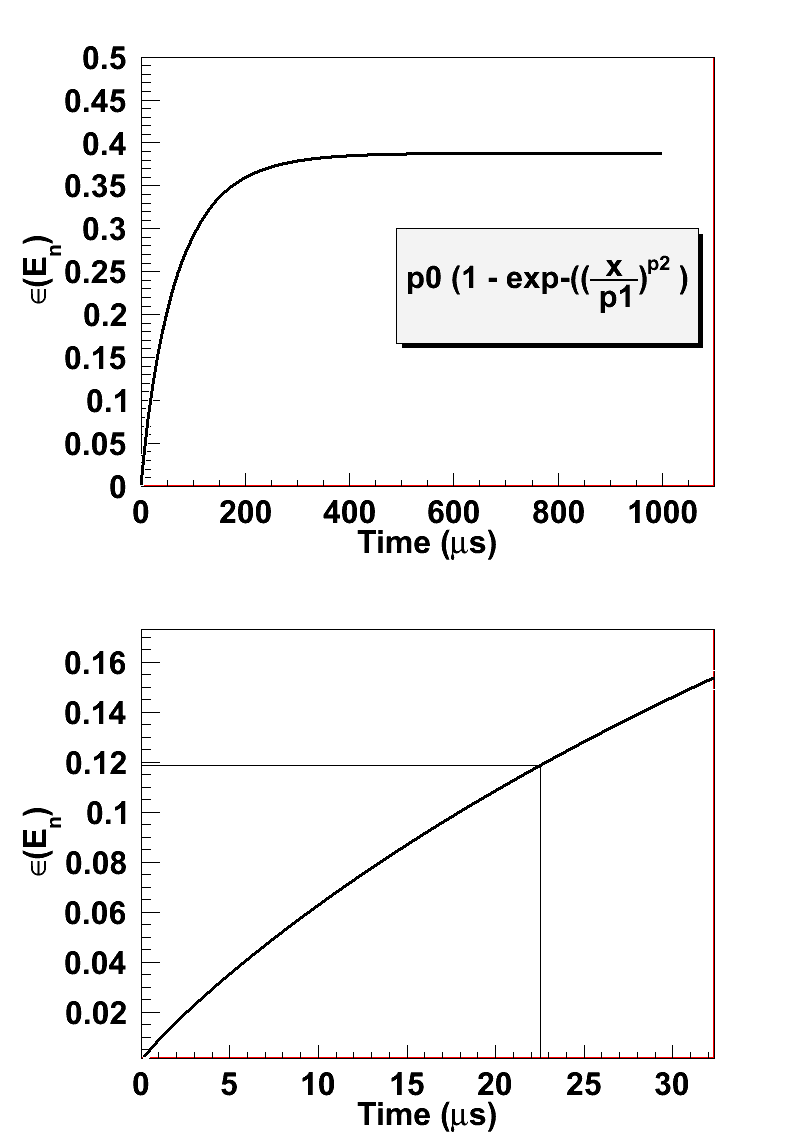}
\caption{\label{Fig:ddneffplot}(Color Online) (Top) A three parameter fit (solid line) describes the simulated efficiency of neutron detection as a function of time in the $^{2}$H($d,n$)$^{3}$He experiment very well.\\(Bottom) The same plot expanded to show t $<$ 35 $\mu$s.  The vertical and horizontal lines identify the experimental window and expected efficiency.}
\end{figure}

The simulation of the measurements made with the $^{2}$H($d,n$)$^{3}$He reaction were made as follows.  
A simulated beam of 2.26 MeV neutrons was emitted from inside an evacuated volume, through an aluminum beam-pipe wall, directed toward the axial center of a set of polyethylene plugs that filled most of the detector cavity (see Fig.~\ref{Fig:ddngeom}). Because the experiment only recorded counts in the neutron detector during a 22.5 $\mu$s wide time window after the associated $^3$He particle was detected, it was necessary to track the neutron detection time in the simulations.  Therefore, a time dependent model for detection of neutrons emitted from the $^{2}$H($d,n$)$^{3}$He reaction was created to compare with experiment.  In this model, only neutrons detected before a user-defined time counted toward efficiency.  A plot of efficiency vs. time was simulated for times between t = 0 and t = 1000 $\mu$s (see Fig.~\ref{Fig:ddneffplot}).  To produce a simulated TAC spectrum, a plot of the slope of $\epsilon$(t) vs. time was generated for comparison with data (see Fig.~\ref{Fig:TACSPEC}). 

The simulated total efficiency for neutrons collected between 0 and 22.5 $\mu$s is 11.9\% in agreement with experiment.  
It is noteworthy, that simulations predict a relatively long time (nearly 500 $\mu$s) before a maximum efficiency of 38.8\% detection is realized for 2.26 MeV neutrons from the $^{2}$H($d,n$)$^{3}$He reaction. 
\subsubsection{\label{sec:Liand2HMOD}$^{7}$Li($p,n$)$^{7}$Be and $^{2}$H($\gamma, n$)$^{1}$H\protect} 

Simulations for the $^{7}$Li($p,n$)$^{7}$Be and $^{2}$H($\gamma, n$)$^{1}$H reactions were carried out in the following way. First, the location of the source was set to match experimental conditions. For a single simulation the source emitted monoenergetic neutrons only between angles $\theta$ and $\theta$ + $d\theta$ with constant emission over $\phi$.  After stepping through all of $\theta$ space, the process was repeated for a new neutron energy.

Ultimately, a three-dimensional plot was constructed with neutron energy on the x-axis, emission angle on the y-axis and detection efficiency on the z-axis (see Sect.~\ref{sec:SUM}).  After choosing an incident particle energy, and inputing expected angular distributions for the neutrons in the center-of-mass (CoM) frame, a second Monte Carlo process produced an average efficiency for the given source conditions.  This process was repeated for several incident particle energies, and the result was a plot of simulated efficiency as a function of incident particle energy.  
\begin{figure}
\includegraphics[width=0.5\textwidth]{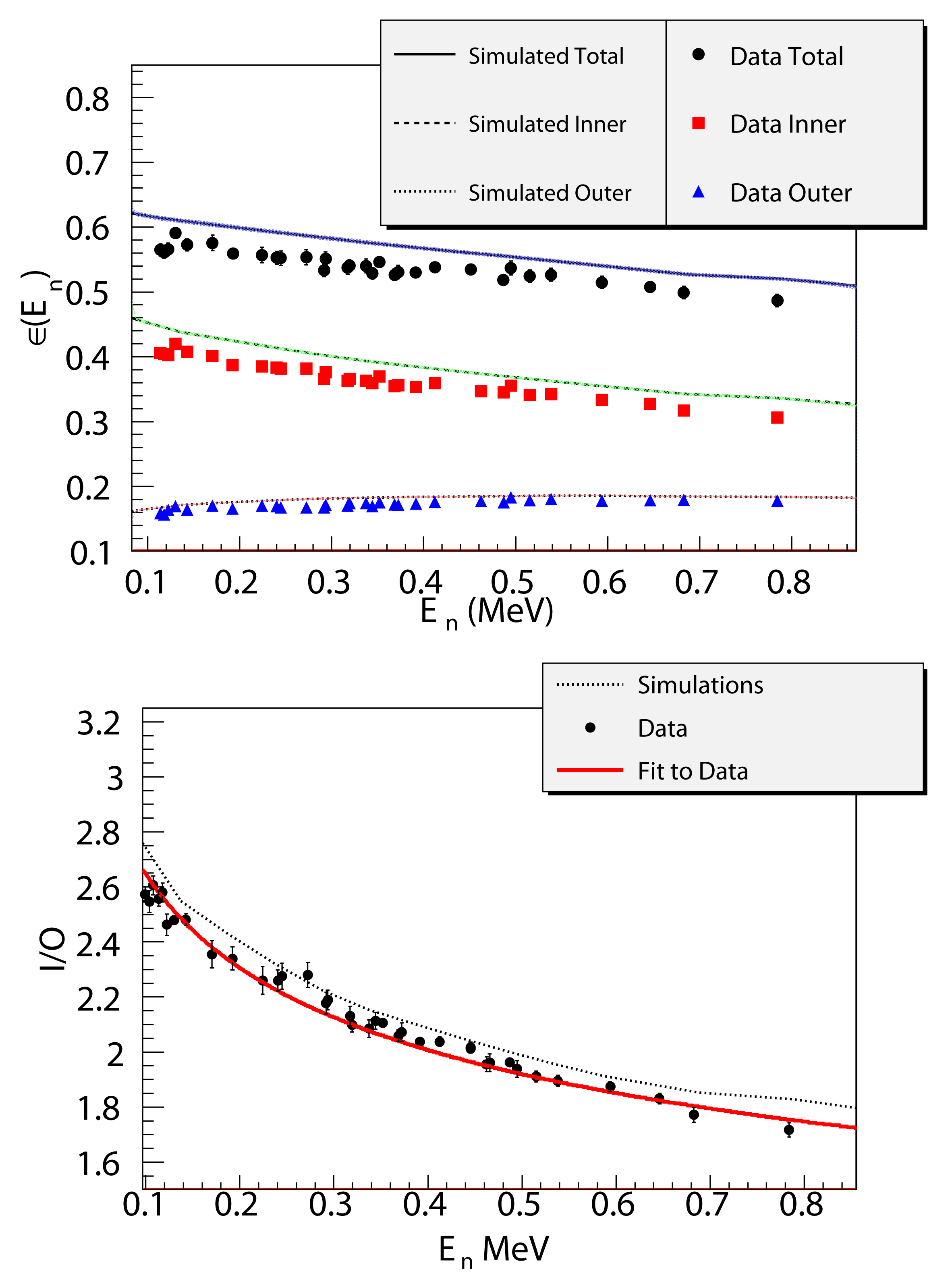}
\caption{\label{Fig:IOE}(Color Online) The efficiency (a) and I/O ratio (b) are shown for neutrons detected during the $^{2}$H($\gamma,n$)$^{1}$H experiment. Statistical error bars are smaller than the data points.}
\end{figure}

\begin{figure}
\includegraphics[width=0.5\textwidth]{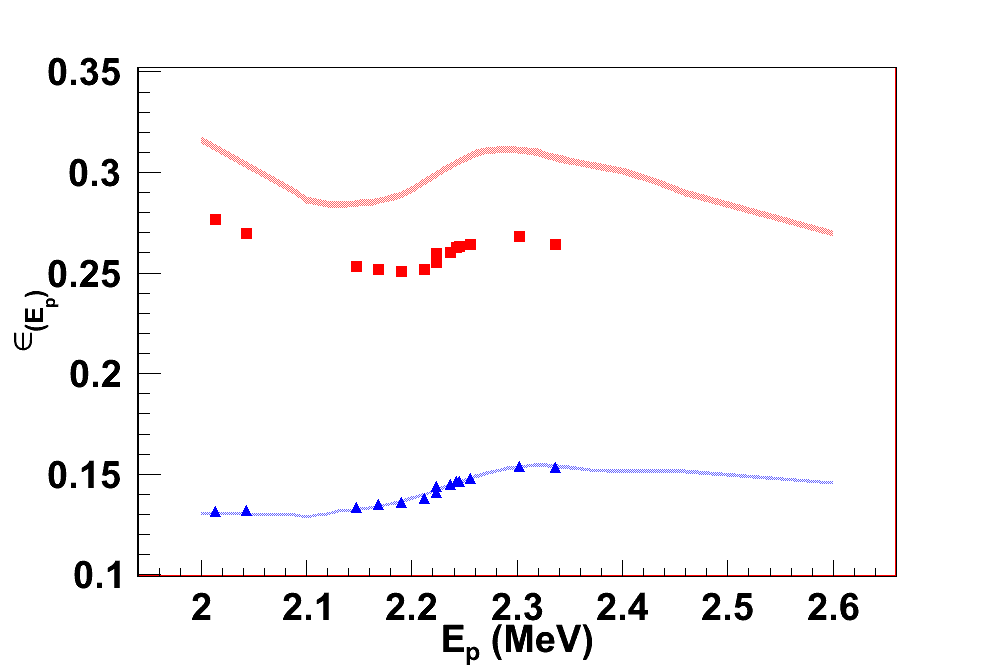}
\caption{\label{Fig:Li7_IO}(Color Online) Absolute efficiency for the inner ring (squares) and outer ring (triangles) of the detector during the $^{7}$Li($p,n$)$^{7}$Be experiment.  Statistical uncertainties are smaller than the data points. Simulations with uncertainties are shown as colored bands.}
\end{figure}

For the $^{7}$Li($p,n$)$^{7}$Be reaction, simulations reproduce very well the shapes of $\epsilon$(E$_{p}$) for both the inner and outer detector rings (see Fig. \ref{Fig:Li7_IO}).  Absolute detection efficiency for the outer ring of detectors is in good agreement with experiment.  A 13\% systematic offset in absolute efficiency is observed for the inner ring of detectors.  Known systematic effects can account for a maximum difference of 6.6\%.  The difficulty with this discrepancy is that it appears to be of a systematic nature, while only affecting the inner ring of the INVS counter.  A missing thermal neutron sink in the model may explain the difference.  An underestimated amount of aluminum in the modeled beam pipe could have inadequately converted neutrons to $\gamma$-rays, leaving an excess of thermal neutrons in the region of the inner ring of the INVS counter.

For the $^{2}$H($\gamma,n$)$^{1}$H experiment plots of $\epsilon$(E$_{n}$) for the inner ring, the outer ring, and the total show the data trends in good agreement with trends predicted by the simulation. 
The data for total detection efficiency are systematically lower than simulation by about 5.9\%.  Data for the inner and outer rings were systematically lower by 6.7\% and 3.7\% respectively. These data provide a benchmark calibration for this INVS counter with regards to its use in future ($\gamma,n$) experiments. 

In the INVS counter a single detected neutron provides no information about the energy of the neutron.  However, the average neutron energy from an ensemble of detected neutrons may be gleaned from the observed proportionality 
\begin{equation}
I/O \propto E_{n}^{-\frac{1}{5}}
\label{eqn:IOpropo}, 
\end{equation}
\noindent which is easily inverted.
The ability to distinguish the signature I/O ratio for E$_{n}$ from the I/O ratio for E$_{n} +$ $\Delta$E$_{n}$ becomes more difficult as E$_{n}$ increases (see Fig. \ref{Fig:IOE}). 

The method described above for evaluating detector efficiency is valid irrespective of the target used.  A deuteron target was chosen because of the precision with which the ($\gamma,n$) cross section is known for this nucleus.   Evaluating the efficiency this way established the energy-dependent response of the detector for future users under certain conditions.  The efficiencies determined here are valid for any ($\gamma,n$) reaction measurement that satisfies the following conditions: (a) the target location was fixed to match simulation; (b) the average energy of the emitted neutrons is known; and (c) the CoM angular distribution is known. 
If backgrounds are low, and the neutron energies are $<$ 2.0 MeV, condition (b) may be relaxed, because the neutron energy information may be obtained from the I/O ratio.  For neutron energies $<$ 500 keV, condition (c) may also be relaxed because the response of the detector is nearly constant with respect to angle of emission.  

\section{\label{sec:SUM}Summary and Conclusions:\protect}

The goal of this work was to characterize precisely the response of the highly efficient INVS counter using multiple neutron sources with a focus on neutrons of energy $<$ 1.0 MeV.  The attention to low energy neutrons was motivated by a need to generate high quality ($\gamma,n$) cross-section data.  
Experiments were carried out and then simulated in detail for comparison.  In all simulations, special attention was given to assure accurate reproduction of experimental conditions. 

\begin{figure}
\includegraphics[width=0.5\textwidth]{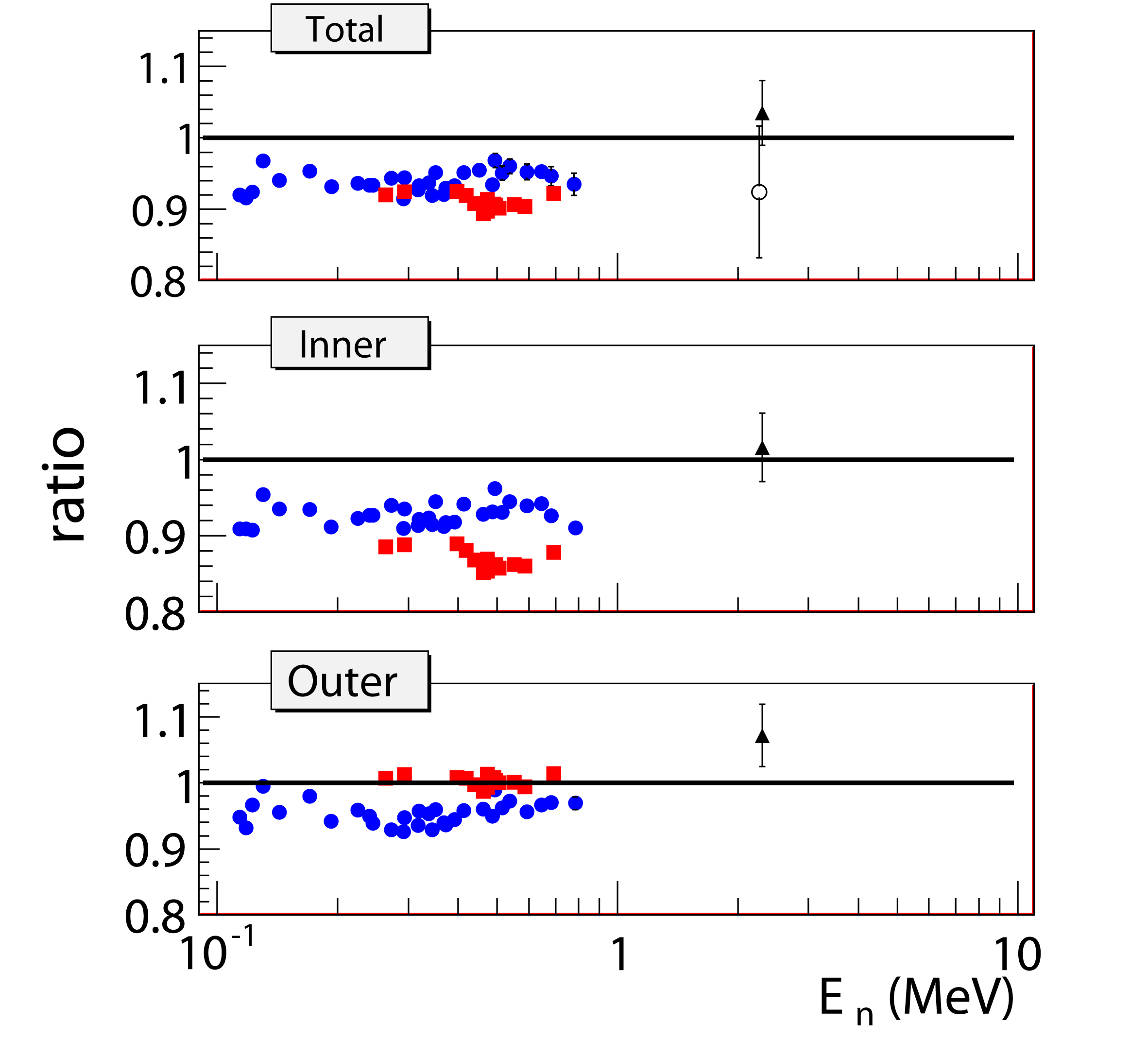}
\caption{\label{Fig:AllPlot}(Color Online) Plot of the ratio of data to simulation for all experiments.  Results for total $\epsilon_{n}$, inner ring $\epsilon_{n}$, and outer ring $\epsilon_{n}$ are shown.  Blue filled circles are $^{2}$H($\gamma,n$)$^{1}$H data; red squares are $^{7}$Li($p,n$)$^{7}$Be data; black triangles are $^{252}$Cf data; open circles are $^{2}$H($d,n$)$^{3}$He data.}
\end{figure}\begin{figure}
\includegraphics[width=0.5\textwidth]{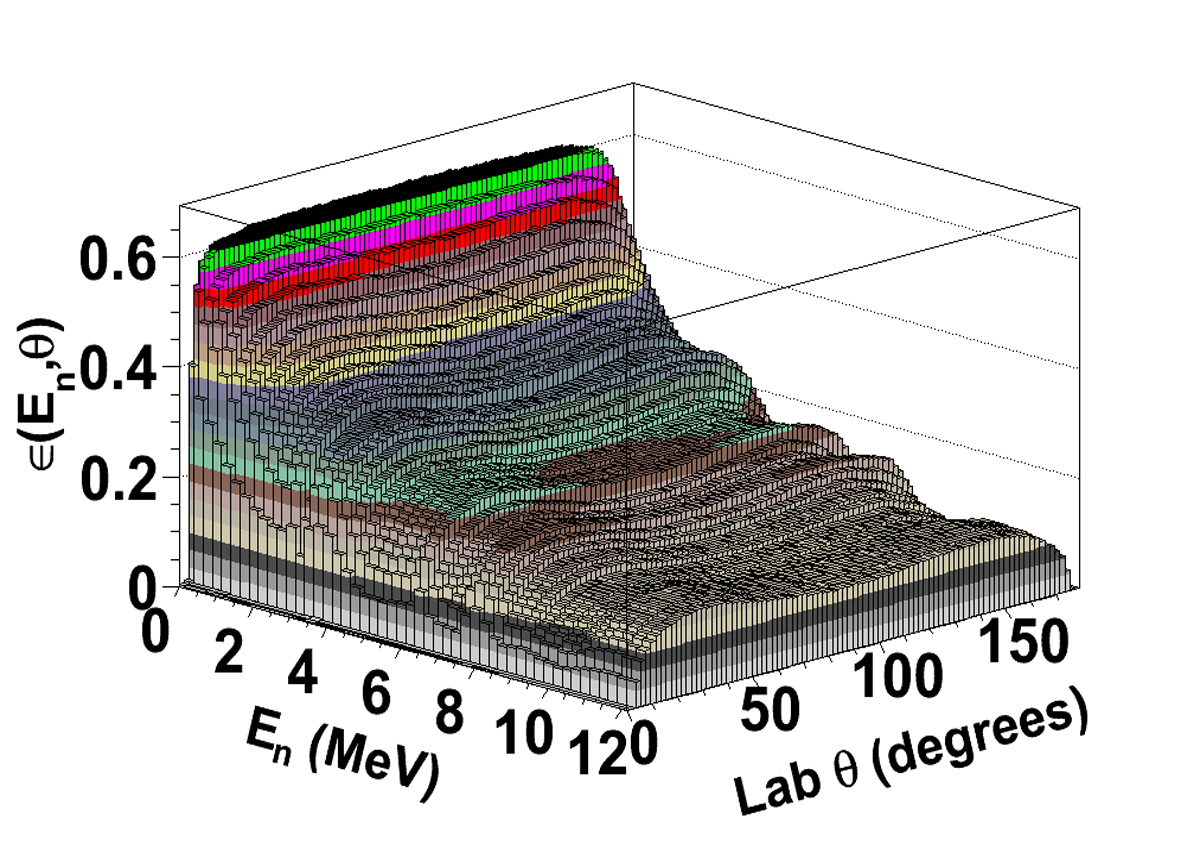}
\caption{\label{Fig:gendet} A three-dimensional plot of efficiency vs E$_{n}$ and $\theta_{lab}$ using a likely arrangement of moderator for ($\gamma,n$) experiments at HI$\gamma$S.  The simulated neutron-emitting target was at the axial and longitudinal center of the detector. The simulated geometry filled the detector cavity with graphite except for a 2.54 cm diameter hole for the target and $\gamma$-ray beam. }
\end{figure}


Figure~\ref{Fig:AllPlot} shows the ratio of experimentally determined efficiency to simulated efficiency vs. average neutron energy for each experiment.  Ratios have been determined for the inner ($I$) and outer($O$) rings separately, as well as for the total ($T$).  

Neutrons from the $^{2}$H($\gamma,n$)$^{1}$H reaction were emitted from within $\pm$ 4 cm from the longitudinal center of the detector with a sin$^{2}$($\theta$) distribution in the CoM frame which is hardly changed when converted to the lab frame because of the relatively small momentum of the incident $\gamma$-ray.  The arrangement of moderating materials was approximatly symmetric. Under these conditions, simulations reproduce measurements to within a normalization of -6.7\% ($I$), -3.7\% ($O$), and -5.9\% ($T$).   These differences are likely the result of effective threshold settings on the signals from the detector tubes.  This feature of the effective threshold settings was not included in the simulations.

Neutrons from the $^{7}$Li($p,n$)$^{7}$Be reaction were emitted far from the longitudinal center of the detector with a Legendre polynomial series distribution in the CoM frame which is significantly changed when converted to the lab frame because of the relatively large momentum of the incident proton.  The arrangement of moderating materials was asymmetric - biased to achieve higher detection efficiency for neutrons emitted near $\theta_{lab}$ close to zero. Under these conditions, absolute detection in $O$ is reproduced very well by simulations; however, absolute detection in $I$ is systematically 13\% smaller than predicted by simulations, resulting in a nearly 10\% systematic difference in $T$.

The difference in the level of agreement between simulations and experiment for $I$ and $O$ for the $^{7}$Li($p,n$)$^{7}$Be reaction is intriguing.  The most likely explanation is that the amount of aluminum in the intervening beam pipe was underestimated in the simulations which caused an excess of thermal neutrons in the vicinity of $I$.
  
The systematic differences between experiment and simulation for $I$, $O$ and $T$ for the $^{2}$H($\gamma,n$)$^{1}$H measurement confirmed the need for a well known, tunable, monoenergetic neutron source.  Reliance on simulations alone would have introduced systematic errors in future ($\gamma,n$) measurements on the order of 6\%.  Using the cross-section of Ref.~\cite{Sch05} as a 1\% standard provided tunable monoenergetic neutron sources with fluxes known to $\pm$ 3\%.  The method for simulating absolute detection efficiencies for the $^{2}$H($\gamma,n$)$^{1}$H reaction was not significantly influenced by the choice of target material.  Thus, using the same techniques, other ($\gamma,n$) cross-sections may be measured with very high accuracy.

For the purpose of planning future ($\gamma,n$) experiments at HI$\gamma$S which seek to use the INVS counter, simulations have been prepared for three likely experimental setups. An analysis program takes as input the geometry, the desired ($\gamma,n$) reaction, the incident particle energy, and the CoM angular distribution of the outgoing neutrons and gives as output the expected detection efficiency and I/O ratio. 
The capabilities offered by the combination of the HI$\gamma$S facility and this precisely characterized INVS counter make possible absolute photonuclear cross-section measurements with high precision.

\begin{acknowledgments}
Work supported in part by USDOE Office of Nuclear Physics Grants DE-FG02-97ER41041 and DE-FG02-97ER41033.  
I would like to acknowledge the staff of the UNC Chapel Hill and TUNL machine shops, and experimental collaborators A. Adekola, G. Rusev, S. Stave, M.W. Ahmed and Y. Wu for their help in completing these experiments.
\end{acknowledgments}

\bibliography{NDP_v2}

\end{document}